# A POSSIBLE MODEL FOR THE MASS SPECTRUM OF ELEMENTARY PARTICLES

**Leonardo Chiatti**
AUSL Medical Physics Laboratory
Via Enrico Fermi 15, 01100 Viterbo (Italy)
fisica1.san@asl.vt.it

**Summary**

A conjecture on the origin of elementary particle masses is discussed, based on the "micro-universe" and quantum state reduction concepts. The reduction of the quantum state of a real particle is understood to take place objectively; in every interaction event with other real particles, the quantum state of the particle is annihilated and a new one is created. In these events, the particle is localised in time and in space, with a maximum precision established by the uncertainty principle. The minimum energy required for this localisation constitutes the chief contribution to the particle's mass. A further contribution is provided by the self-interaction of the particle during localisation, which occurs within a finite duration. Some algorithms for estimating these two contributions are presented and discussed.

## 1. Introduction

A model is conjectured in this article for the calculation of the masses of elementary particles, based on the following assumptions:

1) Let us consider the quantum state associated to a real particle. The discontinuous "jump" of its time evolution (reduction or "collapse" event) is a real, objective physical phenomenon.

2) It occurs every time the particle interacts with systems constituted by other real particles. In these interaction events the quantum state is annihilated, and a new quantum state is created.

3) The objective reduction process indicated under points 1) and 2) has a minimum finite duration, established by the uncertainty principle. Thus, the proper time interval between two successive reduction processes (creation and destruction) involving the same real massive particle must exceed this duration.

4) The spacetime localisation undergone by the particle during reduction requires the exchange, in the interaction vertex, of an energy whose minimum value represents the chief contribution to the mass of the created/destroyed particle.

5) A minimum proper time interval exists between two successive interactions (real or virtual) of a massive particle. This minimum interval constitutes a limit to the precision with which the particle can be localised in time through *successive* interactions. As has been proved in the classical context by Caldirola (1), this interval, multiplied by the speed of light in the vacuum c, constitutes the radius of a De Sitter "micro-universe" associated with the particle; for its physical meaning, please refer to a previous work (2). Empirically, the maximum value of this radius proves to be identical for leptons and for hadrons, and coincides with the electron classical radius.

6) In the case of hadrons, the duration of the reduction process coincides with that of the minimum interval between two interactions, introduced under point 5). In the case of leptons, on the other hand, the reduction process takes place in a proper time interval that is longer than the minimum interval between two interactions.

7) *During* the reduction process, the particle can self-interact through virtual processes. A second contribution to the particle mass derives from this self-interaction. *Between* the two successive reduction processes (creation and annihilation), the particle not interacts with itself and generally it is not localised in space.

The physical background necessary to justify these assumptions will not be discussed in this article; for this, please refer to previous works (2, 3).

This article is organized as follows. In Section 2, the preceding assumptions are applied separately to the case of leptons and to that of hadrons, showing the different physical meaning of the two contributions to mass. Algorithms are also put forward to calculate the second contribution (i.e. the one resulting from self-interaction). In Section 3, the first contribution (resulting from localisation) is discussed, which we have called "skeleton mass"; it is the absolutely dominant one. In sections 4, 5 and 6, some formulas are conjectured for calculating the respective skeleton masses of leptons and hadrons. Section 7 tackles at the issue of how the proposed model could be generalised to excited hadron states, and reference is made to quantitative analyses which have already appeared in the literature.

The exposition style is qualitative and conceptual, as the goal pursued here is to provide a general presentation of this subject. Detailed computations of masses are left for later works.

## 2. Mass and "R" processes

Let us consider the process of reduction ["R process" in Penrose terminology (4, 5)] corresponding to the preparation (creation) or destruction (annihilation) of the quantum state of a single particle. For the purpose of simplicity, we shall concentrate on the case of creation, as annihilation is nothing other than the inverse process. This "process" which, in our analysis, we shall subdivide into distinct steps, actually consists of one single event; the "steps" described are, therefore, actually simultaneous.

We shall firstly assume that one or more pairs of opposites will be generated, among those listed here[1]:

$$e^+e^-, u\bar{u}, d\bar{d}, \mu^+\mu^-, s\bar{s}, c\bar{c}, \tau^+\tau^-, b\bar{b}, t\bar{t} \ .$$

The members of a lepton pair such as $e^+e^-$ can separate and become distinct particles.

---

[1] Neutrinos are not created or annihilated in mass eigenstates, but in oscillating superpositions of these. This model cannot therefore be applied to these. Since the experimental situation of neutrino masses is still being clarified, they will not be discussed further.

This is not permitted, on the other hand, to individual members of a quark pair such as u$\bar{u}$, d$\bar{d}$; one or more quark pairs must be created simultaneously, after which either the individual pairs [e.g. u$\bar{u}$, d$\bar{d}$] or groups of their members [e.g. udd, u$\bar{d}$] separate. We shall postulate that the exiting aggregates are always only mesons, baryons or antibaryons.

Pairs of opposites, in the creation event, act as operators that transform the particles entering into the interaction vertex into those exiting from it. In Fig. 1, these operators have been enclosed in a circle, to distinguish them from entering or exiting "operand" particles.

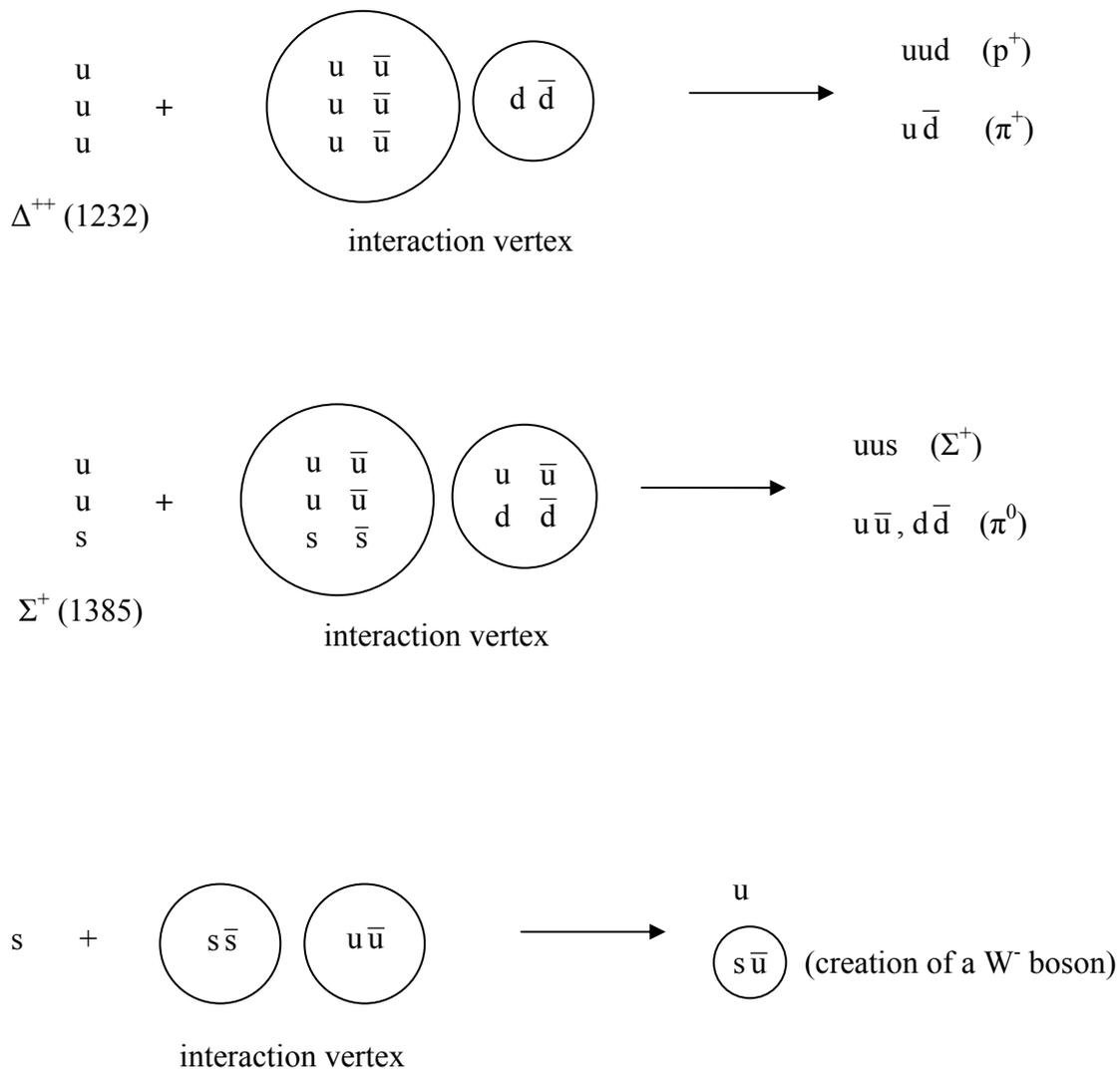

Fig. 1

In this paper, we are interested in determining the mass M($\lambda$) of the particle state $\lambda$, where $\lambda$ is any lepton or hadron. Let us suppose that only the $\lambda$ and $\bar{\lambda}$ states appear at the end of separation; the minimum energy required for these R processes will be M($\lambda\bar{\lambda}$)c$^2$ = 2M($\lambda$)c$^2$.

Based on the above, in the creation of the particle state λ we must distinguish two phases (which are actually simultaneous):

1) the localization, in spacetime, of the lepton or quark group which constitute the particles $\lambda, \bar{\lambda}$;

2) the separation of λ from $\bar{\lambda}$ (which from "virtual" states thus become "real").

In the first phase, the localisation of the particle takes place in a region having minimum space size L. To bring about this localization, an energy ℏc/L is required, by virtue of the uncertainty principle. Dividing this energy by $c^2$, the mass $M_{sk} = \hbar/cL$ is obtained, which shall be called, in this context, the "skeleton" mass of the particle.

At this point, it is necessary to distinguish the case in which λ particle is a lepton from that in which it is a hadron.

1) Leptonic case

At the end of the first phase, the lepton is delocalised over a spatial region whose size is $L = \hbar/M_{sk}c$; however, the radius $r_0 = e^2/M_{sk}c^2 = \alpha L$ of leptonic micro-universe is much smaller than L. Thus, a certain number of copies of the same lepton are present, in a virtual state, over a region whose size is L. Each of these copies derives from the reversal of the lepton motion along the direction of *t*, the time measured in the laboratory (Fig. 2).

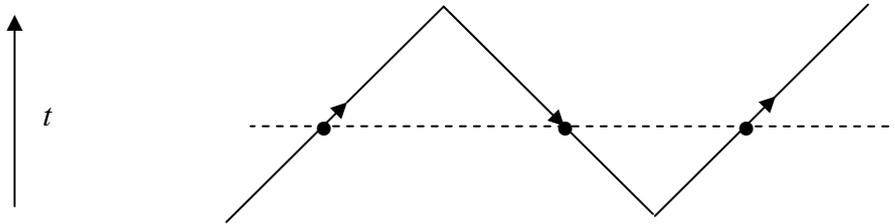

Fig. 2; The different positions of the (virtual) lepton relative to the same time instant *t* constitute the "copies" of the lepton. They are scattered over a spatial region whose size is L. It is to be noted that the total charge of all copies present at time *t* is always the lepton charge, because when it moves with retrograde motion in *t*, the lepton has an opposite charge (3).

These "copies" can interact with each other, exchanging gauge quanta. If we only consider electromagnetic interaction, the interaction energy can be estimated as:

$$\approx e^2/L = e^2/(\hbar/M_{sk}c) = (e^2/\hbar c)M_{sk}c^2 = \alpha\, M_{sk}c^2, \qquad (2.1)$$

in that the total charge of all the "copies" (each of which will have a ±e charge) is equal to the lepton charge (+e or –e).

In the second phase of the creation process, the lepton is transferred from the virtual state to the real state, and once it has reached the real state it is no longer self-interacting. To make this transfer possible, it is necessary to provide the vacuum not only with energy $M_{sk}c^2$, but also with the additional energy $k\alpha M_{sk}c^2$ (where $|k| \approx 1$), equal to the self-interaction energy of the virtual lepton.

This is equivalent to saying that the effective rest mass of the lepton is not $M_{sk}$, but $M_{sk} + k\alpha M_{sk} = M_{sk}(1 + k\alpha)$.

One can suppose that the self-interaction term that must be added to the skeleton mass to obtain the effective mass is expressed by the integral:

$$M_{self} = k\alpha M_{sk} = -\frac{e}{c^2} \int \overline{\Phi} \gamma^\mu A_\mu \Phi \, dV \quad , \tag{2.2}$$

where:

$$\gamma^\mu (\partial_\mu - \frac{ie}{\hbar c} A_\mu) \Phi = -(cM_{sk}/\hbar)\Phi \quad ; \tag{2.3}$$

$$\partial^\nu \partial_\nu A_\mu = -4\pi i e \overline{\Phi} \gamma_\mu \Phi . \tag{2.4}$$

In calculating the integrals that express $M_{self}$ and $A_\mu$, it must be borne in mind that self-interaction is limited (in the particle rest frame of reference) to distances greater than $r_0$; at lower distances, only one lepton copy is present, which certainly cannot interact with itself.

The model outlined here has been proposed, in a different form, by Kawahara (6). Kawahara's non-renormalised mass, obtained by imposing a periodicity condition on the lepton wavefunction, is our skeleton mass, while his renormalized mass is equivalent to our effective mass $M = M_{sk} + M_{self}$. Kawahara, too, executes the integral (2.2) over a region having radius L, assuming that the electric and magnetic electron self-fields are those associated with a point-like source. Instead, he does not assume any lower integration limit, and obtains an expression of the type $M_{self} = k\alpha M_{sk}$, with $k = 1/\pi$, regardless of the lepton taken into consideration.

In our model, on the other hand, the parameter k depends on the leptonic generation index i. The effective masses ratio is thus theoretically expressed by the equation:

$$[m(i)/m(1)]_{eff} = [M_{sk}(i)/M_{sk}(1)] \times [(1 + k_i\alpha)/(1 + k_1\alpha)] . \tag{2.5}$$

The dependence of the k parameter on the generation index i demands an explanation. The number of copies of the lepton contained in the region having spatial size $L = \hbar/M_{sk}c$ and time $\hbar/M_{sk}c^2$ is expressed by the ratio between the time interval $\hbar/M_{sk}c^2$ and the duration of the individual leptonic micro-universe $e^2/M_{sk}c^3$. This ratio is the universal constant $\alpha^{-1} = \hbar c/e^2$, which is independent of the lepton in question. On passing from the i-th to the j-th leptonic generation, the skeleton mass is multiplied by the ratio $z^{-1} = M_{sk}(j)/M_{sk}(i)$, and thus L is multiplied by the ratio z. The radius $r_0 = e^2/M_{sk}c^2$ of the leptonic micro-universe, too, is multiplied by z.

In calculating the self energy of the particle, the integration volume $L^3$ is multiplied by $z^3$ and, since the number of copies located within the region remains unchanged, their spatial density ρ proves to be multiplied by $z^{-3}$.

Let us therefore consider the side Λ of the cube containing only one copy on average; of course, $\rho \Lambda^3 = 1$. Thus, in the passage i → j mentioned above, Λ is multiplied by a factor z and, therefore, the ratio $r_0/\Lambda$ remains unchanged. Consequently, the integral expressing the self energy of the lepton, which is basically a Coulomb integral of the type [ξ(x) = 1 if x belongs to a copy, 0 otherwise]:

$$-e^2 \rho^2 \int \frac{dV \, dV'}{|x - x'|} \xi(x) \, \xi(x') \, ;$$

evaluated in the particle restframe of reference, simply remains multiplied by 1/z. The contribution of self-interaction to the mass $k\alpha M_{sk}(i)$ thus becomes $k\alpha M_{sk}(j)$, but k remains unchanged. Let us suppose, however, that in the virtual cloud in which the lepton of a given generation is dissolved copies of leptons of *other* generations are also present. This is undoubtedly true if we take into account radiative corrections of a higher order. In this case, the size R of copies of a different generation from that of the lepton in question will not be multiplied by z; thus the ratio $R/\Lambda$, with regard to these copies, will not remain unchanged. Thus the Coulomb integral will still be approximately scaled by a $z^{-1}$ factor, yet *not exactly*. The deviation will depend on the structure of the virtual dissociation of the lepton in question, and this, in turn, will depend on the generation to which it belongs. Thus the parameter k will depend on the generation index i.

2) Hadronic case

At the end of the first phase, the hadron is localised in a spatial region having minimum size $L = \hbar/M_{sk}c$, while the radius of the hadron micro-universe $r_0 = \hbar/M_{sk}c$ is equal to L. Consequently, there are no multiple copies of the hadron spread over a spatial volume, but only one copy. One cannot therefore apply the reasoning carried out for leptons to hadrons.

Each hadron is made up of two or three quarks, according to whether it is a meson or a baryon/antibaryon. These quarks can interact with each other or self-interact, exchanging gauge quanta. Though these interactions connect events separated by a spatial interval which is less than $r_0$, or by a time interval which is less than $r_0/c$, this does not contradict the statement expressed under point 5) of the Introduction, because these interactions remain within the copy; they do not couple different copies.

The $M_{self}c^2$ energy is identified here with the mutual interaction and self-interaction energy of quarks within the hadron. For the same reasons seen in the leptonic case, it determines a value contribution $M_{self}$ to the hadron rest mass. This is therefore expressed, similarly to the leptonic case, by the sum of this contribution and of the skeleton mass: $M = M_{sk} + M_{self}$. In the hadronic case, as in the leptonic case, only the skeleton mass is related to the size of the particle micro-universe.

Calculation of the energy $M_{self}c^2$ requires some cautions, as the hadron wave equation is formulated on the De Sitter micro-universe having radius L; one has (see Appendix II):

$$i\hbar \partial_t \Psi(\vec{x}_1,...,\vec{x}_j,...,\vec{x}_N, t) = \sum_{j=1}^{N} \left[ -\vec{\alpha} \cdot \left( \vec{\partial} - q\vec{A} - q_s \hat{\vec{A}} \right) + i\hbar q A_0 + i\hbar q \hat{A}_0 \right]_j \Psi(\vec{x}_1,...,\vec{x}_j,...,\vec{x}_N, t)$$

$$+ M_{sk} \Psi(\vec{x}_1,...,\vec{x}_j,...,\vec{x}_N, t)$$

(2.6)

This equation must be solved for stationary states relating to values of N equal to 2 (mesons) or 3 (baryons/antibaryons). The skeleton mass $M_{sk}$ which appears in equation (2.6) is that of the *entire* hadron.

The confinement of quarks within the radius L of the micro-universe is ensured by the projective deformation of the α matrices. A further confinement of quarks within the distance $\hbar c/E_0 \geq L$, where $E_0 = 70$ MeV, is ensured, on the other hand, by the pattern of the SU(3) $\hat{A}_\mu$ potentials as a function of distance; they diverge at the distance $\hbar c/E_0$. Locally, quarks, apart from electroweak interaction, are free. Similarly to the leptonic case, one has (see Appendix II for details):

$$M_{self} = -\frac{1}{2c^2} \int \overline{\Psi} q_j \Gamma^{\mu j} A_{\mu j} \Psi \, dV \qquad (2.7)$$

where the sum on the quark index j includes both the SU(3) and the SU(2)xU(1) sectors of the interaction.

The energies $c^2 M_{self}$ can be both real and complex; furthermore, the real part can be both positive and negative; nevertheless, the hadron components (quarks and gluons) are in any case confined within the hadron micro-universe. As a result, the positive energies do not correspond to diffusion states. An imaginary finite part implies a finite lifetime for the hadron state in question.

If, in equation (2.6), all the internal forces are cancelled out and a homogenous wavefunction on the micro-universe is assumed, the sum cancels out and only the skeleton term remains, as should be the case. It is clear, on visually inspecting equation (2.6), that individual quark masses do not exist - only a global skeleton mass $M_{sk}$ exists. In retrospect, this is perhaps a justification of the success of methods based on unitary symmetry in interpreting hadronic flavour wavefunctions as superpositions of the $(u\bar{u})$-$(d\bar{d})$ etc. form. In these superpositions, the flavour of the individual quark is not defined, and this would not be possible if different flavour states corresponded to different mass values.

## 3. Skeleton masses

To calculate the mass of a particle, one must therefore firstly determine its skeleton mass and from this point on we shall speak exclusively of skeleton masses. We must bear in mind that these are defined directly by the process of the particle emergence from the vacuum (creation) or by the inverse process of annihilation. We can assume the skeleton mass spectrum to be a fundamental property of the vacuum, as is the case in classical physics for the fundamental constants $\varepsilon_0$ and c. In a certain sense, in this perspective the skeleton mass is a substitute for the Higgs field.

We search for equations giving the skeleton mass of leptons and hadrons, respectively. Due to the reasons explained previously, these formulas cannot be deduced from dynamics but will have to be inferred phenomenologically by the regularities in the spectrum of *effective* particle masses[2].

As regards charged leptons, the skeleton mass will have to be expressed as a function of the leptonic generation (family) index. The relevant mass formula must therefore be inferred by examining the dependence of effective masses on the generation index, bearing in mind that the contribution of self-interaction to the effective mass - which is basically electromagnetic here - is approximately 1%, as described in the previous section.

As regards hadrons, some considerations are called for. Equation (2.6) tells us that individual quarks have no mass; thus, with respect to the internal dynamics of the hadron, all quark flavours are equivalent. The contribution of electroweak interaction constitutes an exception, as different charges are associated with different flavours. This contribution basically gives rise to the separation of the members of an SU(2) isospin multiplet.

In other terms, equation (2.6) is substantially SU(N)-invariant, except for the splitting of the SU(2) multiplets. The greater part of the breaking of the $SU(N)_{flavour}$ symmetry is thus originated by the dependence of the skeleton mass $M_{sk}$ on the isospin multiplet to which the hadron belongs. To each isospin multiplet there correspondes a value of the hadronic micro-universe radius, and this determines the splitting among the average masses of different isospin multiplets. Thus, the formula which expresses the skeleton mass of hadrons must be sought, on a phenomenological basis, by examining the dependence of the effective masses on the isospin multiplet to which the hadron belongs.

---

[2] Probably, skeleton mass formulas can be justified in the context of a "creation algebra" linking such aspects as zitterbewegung, the structure of creation-annihilation operators, spacetime geometry and standard model. First attempts along this direction are represented by refs. (9), (22), (23).

The mass formulas will actually have to be determined for the $\lambda\bar{\lambda}$ pair, having the quantum numbers of the vacuum, where $\lambda$ is the generic lepton or *hadron*. Despite the fact that leptonic and quark flavours appear in increasing effective mass order, according to the sequence:

| | |
|---|---|
| $e^+e^-$ | 1.022 MeV |
| $\pi^0$ ($u\bar{u}$, $d\bar{d}$) | 134.98 MeV |
| $\mu^+\mu^-$ | 211 MeV |
| $\varphi$ ($s\bar{s}$) | 1019.4 MeV |
| $\eta_c$ ($c\bar{c}$) | 2979.7 MeV |
| $\tau^+\tau^-$ | 3556 MeV |
| $Y$ ($b\bar{b}$) | 9460.3 MeV |
| $t\bar{t}$ ? | ? |

which clearly has the structure:

lepton (charged), quark, quark, lepton (charged), quark, quark, lepton (charged), quark, quark,

it must be borne in mind, however, that actually, in hadron creation processes, hadrons are generated, and not individual quarks of opposite flavour. In such processes, after a number of quark pairs $q\bar{q}$ are generated (where q is a given flavour u, d, s, etc.) the quark aggregates separate, which constitute the hadron $\lambda$ and the anti-hadron $\bar{\lambda}$, respectively. The mass to be taken into consideration is actually the mass of these aggregates.
Nevertheless, the content of the following Sections 4 and 5 can be considered also in reference to the sequence described above.

**4. Leptonic skeleton spectrum**

The formula suggested for the skeleton mass m(i) of the i-th generation charged lepton is as follows[3]:

$$m(i > 1)/m(1) = (3/2)(i-1)^4 \alpha^{-1} [1 + (i-1)^3 \alpha] . \qquad (4.1)$$

---
[3] We propose here a formula that is similar to the well-known Barut formula (7, 8), but slightly different. Both formulas are easily justified within the context described in ref. (9), Chapter XII. For example, N(i-1) = (i-1)³ is the number of equivalent paths on the (i-1)-th generation lepton graph, which connect a "tower" to itself.

One has:

m(1)/m(1) = 1

m(2)/m(1) = 207.05385

m(3)/m(1) = 3480.8616

**5. The sequence of leptonic and quark flavours**

It is assumed that the pion skeleton mass is expressed as a function of that of the electron by the relation $M_\pi = \alpha^{-1} M(e^+e^-) = 140$ MeV.
As regards lightest mesons having a $q\bar{q}$ structure (where q is a given flavour u, d, s, etc.), selected regardless of their spin, these form the "tower" shown in Fig.3.

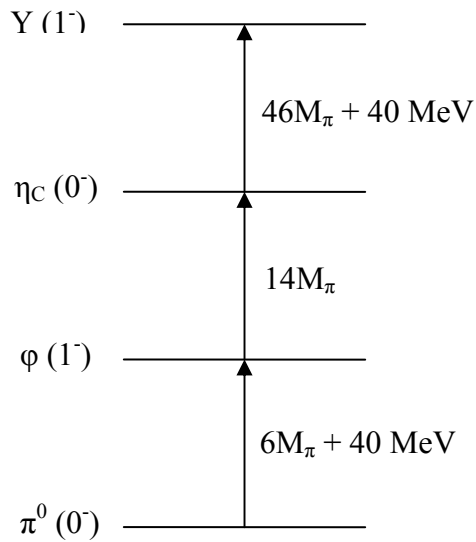

Fig. 3

By indicating with $\Delta M_j$ the j-th difference in mass in ascending order, and letting $\Delta m_0 = 5 M_\pi$, the structure of the tower is summarised in the formula:

$$\Delta m_{j+1} = (j+1) \times \Delta m_j + 2^j M_\pi ;$$

$$\Delta M_{j+1} = \Delta m_{j+1} + Q_{j+1} .$$
(5.1)

where $Q_j = 40$ MeV if the j-th transition links a pseudoscalar state to a pseudovectorial one, 0 in the opposite case. The following results, which are practically exact, are obtained:

$\pi$ (u$\bar{u}$, d$\bar{d}$) : 140 MeV (assumed)

$\varphi$ (s$\bar{s}$) : 1020 MeV

$\eta_C$ (c$\bar{\text{c}}$) : 2980 MeV

Y (b$\bar{\text{b}}$) : 9460 MeV

Equation (5.1) provides mass values that are very close to the measured effective masses and, together with equation (4.1), describes very well the sequence illustrated in Section 3. However, equation (5.1) has not been derived with the criteria set forth in Section 3, i.e. by analysing the dependence of effective masses on isospin multiplets; thus, the masses obtained with this formula cannot be considered as skeleton values. Equation (6.1) below constitutes a possible general formula for hadronic skeleton masses.

## 6. Hadronic skeleton spectrum

Let us consider the pair, having vacuum quantum numbers, $\lambda$-$\bar{\lambda}$ where $\lambda$ is the generic hadron. We assume the following general mass formula for it:

$$M(\lambda\bar{\lambda}) = \left[(2v_i +1)(2C_i +1+4m_i)+(2l_{i,n} +1)h_{i,n} +4k_{i,n}\right]\alpha^{-1} M_{electron}c^2 \quad . \tag{6.1}$$

In this formula, the index $i$ relates to a set of SU(3)$_{flavour}$ multiplets, while the index $n$ identifies, within this set, the isospin multiplet to which the particle $\lambda$ belongs. The following rules are assumed, on an empirical basis:

$C_i = v_i + m_i$ for all mesons with $i \neq 0$ (6.2a)

$C_i = v_i + m_i + 1$ for all baryons (6.2b)

$C_i = i$ for all hadrons composed solely of u, d, s quarks or related antiquarks (6.2c)

$h_{i,n} = 0,1; \quad k_{i,n} = l_{i,n}, l_{i,n}+ 1$ (6.2d)

The relation between indices $l$ and $k$ is summarised in Table I (Appendix I). The coefficients of the mass formula (6.1) are set forth in Table II (App. I), for the different groups of particles in ascending order of mass.
The results obtained with equation (6.1) are summarised in Table III (App. I), where the masses are expressed in $\alpha^{-1}M(e^+e^-)c^2/2 = 70$ MeV units.

Some remarks on Tables II, III.

1) There is a clear continuity between $\eta$ and $\rho$ mesons, as both have one excitation quantum with a value of 13; on the other hand, there is a leap in $C_i$, which goes from 0 to 1. It is an interesting fact that this passage occurs at the level of really neutral mesons placed at the centre of their respective octets.
    $\omega$(782) has the same numbers as the multiplet $\rho$; mesons $a_0$ and $f_0$ occupy the same position as $\eta$'.

2) As regards Y(b$\bar{\text{b}}$), with a mass of 9460 MeV (= 1/2 * 270.28 in our units) it may be that: $v_i = 3$; $C_i = 6$; $m_i = 6$; $h_{i,n} = 1$; $l_{i,n} = 1$; $k_{i,n} = 2$. Alternatively, it may be that: $v_i = 3$; $C_i = 6$; $m_i = 3$; $h_{i,n} = 1$; $l_{i,n} = 15$; $k_{i,n} = 16$. This second possibility satisfies the empirical rules listed above. In both cases M(YY) = 270. The percentage difference between this value and the experimental one is 0.1%.

In order to identify coefficient formation rules, it is advisable to subdivide the data in Table II by spin multiplets. Thus we have, first of all, pseudoscalar and pseudovectorial mesons, set forth in Tables IV(a), IV(b), respectively (App. I). Baryons having spins of ½, 3/2 are grouped in Tables V and VI (App. I), respectively.
The conjectured rules of formation based on the data set forth in these tables are explained in the related notes.

**7. Hadronic spectrum. Excited states.**

The hadrons considered in the previous sections roughly correspond to the ground states of dynamics described by equation (2.6). The higher-energy stationary or quasi-stationary states correspond to the excited states of the hadron described by the ground state; the radius of the micro-universe L, and therefore the skeleton mass of all these states, are the same as those of the ground state. By very roughly approximating the hadronic micro-universe with a spherical potential well of infinite height and width L, the localisation of massless quarks within this well leads to energy levels that are integer multiples of $\hbar c/L$; i.e. [equation (6.1)] integer multiples of 70 MeV.
In addition, bound states or resonances can be generated as an effect of the interaction of more hadrons. These states are necessarily confined within a region whose size is equal to the range of the strong force, i.e. $\approx d_0 = e^2/mc^2$, where m is the mass of the electron. In other terms, rather than separate hadrons we shall actually have agglomerates of massless quarks confined within a micro-universe whose size is $d_0$. This is the maximum size possible for a hadronic (and leptonic, as well) micro-universe; the energy required to localise a massless quark in this region is roughly $\hbar c/d_0 = 70$ MeV, so that the rest energy of the state will be close to an integer multiple of this value.

In some interesting works (10, 11, 12, 13, 14, 15, 16), Palazzi has examined the families of excited states of different mesons: η, π, b, ρ, a, K, K*, h, ω, φ, f, $D^+$, $D_s$, $η_c$, ψ, $χ_c$, B, $B_s$, Y, $θ^+$. His analysis conclusively proves that the mass levels in the same family are separated by integer multiples of a mass quantum of approximatively 70 MeV. Palazzi bases his fitting on a variable mass quantum of about 35 MeV, and then studies the dependence of this quantum on the spin and on the hadronic family. As a basis for explanation of the observed phenomena, he chooses a composite hadron model, according to which the number of partons is given by the number of mass quanta contained in each hadron. This theoretical choice, however, departs from the quark model and from conventional hadronic physics, with which we, instead, believe we ought to comply. The above discussion can be an alternative to Palazzi's proposal.

In some recent works (17,18), Sidhart has proposed the following unified formula for hadronic masses:

$$M = m (n + ½) \qquad (7.1)$$

in which M is the mass of the hadron expressed in pionic mass units, and *m*, *n* are integer numbers. This formula approximates measured hadronic masses with an accuracy that is comparable to that obtainable with equation (6.1), whose structure it maintains. However, the factors *m*, *n* do not always can be identified with the coefficients of equation (6.1), as is clear in the case of η(547), for which (*m*,*n*) = (8,0). Furthermore, they cannot be clearly correlated with the internal quantum numbers of the particle. Sidhart interprets equation (7.1) as an effect of the presence, within the hadron, of *m* independent harmonic oscillators having zero-point energy equal to half of the pionic mass. This interpretation, however, appears hardly credible, as these oscillators, of which there can even be a high number [8 for η(547) and 16 for $\chi_{b_0}$(1P)] would then all have to be excited at the same, *n*-th level. Assuming that the harmonic oscillator approximation applies at least for the first *n* + 1 energy levels and that the independent oscillators are *m*, the a priori probability of this result should be less than $1/(n+1)^m$. For $\Xi_c^+$, for which $(m,n) = (2,9)$, it should be less than $10^{-2}$.
Furthermore, the approximate quantisation of the masses, with a mass quantum of approximately half a pionic mass, is a phenomenon which also involves charged leptons; for these particles, one cannot suppose a substructure in terms of elementary oscillators related to the shape of the interquark potential, as proposed by Sidhart (18). In our opinion, the validity of equation (7.1) lies upon the physics discussed in this article. Equation (6.1) is probably an improved version of equation (7.1).

## 8. Conclusions

The point of view suggested in this article is consistent with a relational approach to elementary particles: each particle - or, if one prefers, each quantum state of an individual particle - is a link between two events constituted by its creation and by its annihilation. These events constitute all that physical reality offers for exploration. The particle is not understood as a persistent object in motion, with the passage of time, in a "external" environment; thus, in particular, any conception of mass as a property of some "substance" that constitutes the particle is avoided. Rather, the concept of mass is related to the particle localisation in spacetime which occurs when it is created or annihilated as a *real* particle. That is, we are dealing with events corresponding to effective quantum leaps (with quantum state reduction or "collapse") and virtual interactions are not taken into consideration.
The essence of the proposal is that the mass is the sum of two contributions: a dominant one (skeleton mass), linked to the size of the particle micro-universe; the other is associated with particle self-interaction during the localisation process. Some algorithms are proposed to estimate these two contributions [equations (2.2)-(2.7); Section 4, 5, 6].
Though the skeleton mass generation pattern presents a very similar structure to that of the filling of the levels of a bound system, it is probable that one is actually dealing with a sort of quantisation of the radius of the particle micro-universe, valid both for leptons and for hadrons. This would be a feature of the localisation process, and this is consistent with the general sense of this proposal.
The model appears consistent both with Kawahara's calculations on electrons, which follow a similar line of reasoning, and with the analyses conducted by Palazzi and by Sidhart on the mass spectrum of baryons and mesons. Since a detailed verification of this conjecture requires that the proposed formalism be applied to individual particles, it must be left for later works.

# Appendix I - Tables

## Table I

| $l$ | $2l + 1$ | $k$ | $(2l + 1) + 4k$ |
|---|---|---|---|
| 0 | 1 | 0, 1 | 1, 5 |
| 1 | 3 | 1, 2 | 7, 11 |
| 2 | 5 | 2, 3 | 13, 17 |
| 3 | 7 | 3, 4 | 19, 23 |

## Table II

| | $v_i$ | $C_i$ | $m_i$ | $h, l, k$ |
|---|---|---|---|---|
| $i=0$<br>$0^-$ mesons | 1 | 0 | 0 | 1; 0; 0 pion<br>1; 1; 2<br>1; 2; 2 |
| $i=1$<br>$1^-$ mesons | 1 | 1 | 0 | 1; 2; 2<br>1; 2; 3<br>1; 3; 3 |
| $i=2$<br>½-baryons | 1 | 2 | 0 | 1; 1; 2<br>1; 2; 3<br>1; 3; 3<br>1; 3; 4 |
| $i=3$<br>3/2-baryons | 2 | 3 | 0 | 0; 0; 0<br>0; 0-1; 1<br>0; 1-2; 2<br>0; 2; 3 |
| $i=4$<br>$0^-$-charmed (c=1) mesons | 2 | 3 | 1 | 0; 0; 0 |
| $i=5$<br>$1^-$-charmed (c=1) mesons | 2 | 3 | 1 | 0; 1; 1 |
| $i=6$<br>½ -charmed (c=1) baryons | 2 | 4 | 1 | 0; 0; 0<br>1; 0; 1<br>1; 1; 1<br>1; 1; 2 |
| $i=7$<br>3/2-charmed (c=1) baryons | 2 | 4 | 1 | 1; 1; 1<br>1; 1; 2 |
| $i=8$<br>$0^-$-$1^-$ $c\bar{c}$ mesons | 2 | 4 | 2 | 0; 0; 0<br>0; 1; 1 |
| $i=9$<br>$0^-$ b= ±1, s=0 mesons | 3 | 5 | 2 | 1; 2; 3 |
| $i=10$<br>b= ±1, s= ±1, c = 0 mesons | 3 | 5 | 2 | 1; 3; 3 |
| $i=11$<br>$1^-$ b= ±1, s=0 mesons | 3 | 5 | 2 | 1; 3; 4 |
| $i=12$<br>½ - (b= -1) baryons | 3 | 6 | 2 | 1; 2; 2 |
| $i=$ ?<br>$1^-$ $b\bar{b}$ mesons | 3 | 6 | 3 | 1; 15; 16 |

**Table III**

<table>
<tr><td colspan="5" align="center">$i=0$    $0^-$- mesons</td></tr>
<tr><td>$n=0$</td><td>$\pi$</td><td>$M(\pi\pi) = 3\times 1 + 1 = 4$</td><td>Mexp = 3.94</td><td>Diff = -1.5 %</td></tr>
<tr><td>$n=0$</td><td>K</td><td>$M(KK) = 3\times 1 + 11 = 14$</td><td>Mexp = 14.16</td><td>Diff = 1.1 %</td></tr>
<tr><td>$n=1$</td><td>$\eta$</td><td>$M(\eta\eta) = 3\times 1 + 13 = 16$</td><td>Mexp = 15.64</td><td>Diff = -2.2 %</td></tr>
<tr><td>$n=2$</td><td>$\eta'$</td><td>$M(\eta'\eta') = 3\times 1 + 11 + 13 = 27$</td><td>Mexp = 27.36</td><td>Diff = 1.3 %</td></tr>
<tr><td colspan="5" align="center">$i=1$    $1^-$ -mesons</td></tr>
<tr><td>$n=0$</td><td>$\rho$</td><td>$M(\rho\rho) = 3\times 3 + 13 = 22$</td><td>Mexp = 22.03</td><td>Diff = 0.14 %</td></tr>
<tr><td>$n=1$</td><td>K*</td><td>$M(K^*K^*) = 3\times 3 + 17 = 26$</td><td>Mexp = 25.52</td><td>Diff = -1.8 %</td></tr>
<tr><td>$n=2$</td><td>$\varphi$</td><td>$M(\varphi\varphi) = 3\times 3 + 19 = 28$</td><td>Mexp = 29.13</td><td>Diff = 4.0 %</td></tr>
<tr><td colspan="5" align="center">$i=2$    ½-baryons</td></tr>
<tr><td>$n=0$</td><td>N</td><td>$M(NN) = 3\times 5 + 11 = 26$</td><td>Mexp = 26.83</td><td>Diff = 3.2 %</td></tr>
<tr><td>$n=1$</td><td>$\Lambda$</td><td>$M(\Lambda\Lambda) = 3\times 5 + 17 = 32$</td><td>Mexp = 31.88</td><td>Diff = -0.4 %</td></tr>
<tr><td>$n=2$</td><td>$\Sigma$</td><td>$M(\Sigma\Sigma) = 3\times 5 + 19 = 34$</td><td>Mexp = 34.09</td><td>Diff = 0.3 %</td></tr>
<tr><td>$n=3$</td><td>$\Xi$</td><td>$M(\Xi\Xi) = 3\times 5 + 23 = 38$</td><td>Mexp = 37.66</td><td>Diff = -0.9 %</td></tr>
<tr><td colspan="5" align="center">$i=3$    3/2-baryons</td></tr>
<tr><td>$n=0$</td><td>$\Delta$</td><td>$M(\Delta\Delta) = 5\times 7 = 35$</td><td>Mexp = 35.2</td><td>Diff = 0.6 %</td></tr>
<tr><td>$n=1$</td><td>$\Sigma^*$</td><td>$M(\Sigma^*\Sigma^*) = 35 + 4 = 39$</td><td>Mexp = 39.56</td><td>Diff = 1.4 %</td></tr>
<tr><td>$n=2$</td><td>$\Xi^*$</td><td>$M(\Xi^*\Xi^*) = 35 + 8 = 43$</td><td>Mexp = 43.81</td><td>Diff = 1.9 %</td></tr>
<tr><td>$n=3$</td><td>$\Omega$</td><td>$M(\Omega\Omega) = 35 + 12 = 47$</td><td>Mexp = 47.78</td><td>Diff = 1.6 %</td></tr>
<tr><td colspan="5" align="center">$i=4$    $0^-$ charmed (c=1) mesons</td></tr>
<tr><td>$n=0$</td><td>$D^0, D^+$</td><td>$M(DD) = 5\times 11 = 55$</td><td>Mexp = 53.36</td><td>Diff = -3.0 %</td></tr>
<tr><td>$n=1$</td><td>$F^+$</td><td>$M(FF) = 5\times 11 = 55$</td><td>Mexp = 56.24</td><td>Diff = 2.2 %</td></tr>
<tr><td colspan="5" align="center">$i=5$    $1^-$ charmed (c=1) mesons</td></tr>
<tr><td>$n=0$</td><td>$D^{0*}, D^{+*}$</td><td>$M(D^*D^*) = 5\times 11 + 4 = 59$</td><td>Mexp = 57.40</td><td>Diff = -2.7 %</td></tr>
<tr><td>$n=1$</td><td>$F^{*+}$</td><td>$M(FF) = 5\times 11 + 4 = 59$</td><td>Mexp = 60.35</td><td>Diff = 2.3 %</td></tr>
<tr><td colspan="5" align="center">$i=6$    1/2-charmed (c=1) baryons</td></tr>
<tr><td>$n=0$</td><td>$\Lambda_{+c}$</td><td>$M(\Lambda\Lambda) = 5\times 13 = 65$</td><td>Mexp = 65.28</td><td>Diff = 0.43 %</td></tr>
<tr><td>$n=1$</td><td>$\Xi_{oc}, \Xi_c^+$</td><td>$M(\Xi\Xi) = 65 + 5 = 70$</td><td>Mexp = 70.54</td><td>Diff = 0.77 %</td></tr>
<tr><td>$n=2$</td><td>$\Sigma_c^{0,+,++}$</td><td>$M(\Sigma\Sigma) = 65 + 7 = 72$</td><td>Mexp = 72.06</td><td>Diff = 0.00 %</td></tr>
<tr><td>$n=3$</td><td>$\Omega_{oc}$</td><td>$M(\Omega\Omega) = 65 + 11 = 76$</td><td>Mexp = 77.07</td><td>Diff = 2.2 %</td></tr>
<tr><td colspan="5" align="center">$i=7$    3/2-charmed (c=1) baryons</td></tr>
<tr><td>$n=0$</td><td>$\Sigma_c^{*\,0,+,++}$</td><td>$M(\Sigma^*\Sigma^*) = 5\times 13 + 7 = 72$</td><td>Mexp = 71.93</td><td>Diff = 0.00 %</td></tr>
<tr><td>$n=1$</td><td>$\Xi_{0c}^*, \Xi_c^{*+}$</td><td>$M(\Xi\Xi) = 65 + 11 = 76$</td><td>Mexp = 73.61</td><td>Diff = -3.10 %</td></tr>
<tr><td colspan="5" align="center">$i=8$    $0^-$-$1^-$ $c\bar{c}$ mesons</td></tr>
<tr><td>$n=0$</td><td>$\eta_c$</td><td>$M(\eta\eta) = 5\times 17 = 85$</td><td>Mexp = 85.13</td><td>Diff = 0.15 %</td></tr>
<tr><td>$n=1$</td><td>$J/\psi$</td><td>$M(JJ) = 85 + 4 = 89$</td><td>Mexp = 88.48</td><td>Diff = -0.58 %</td></tr>
</table>

| | | i=9  0⁻ b= ±1, s=0 mesons | | |
|---|---|---|---|---|
| *n=0* | B±, B⁰ | M(BB) = 7x19 + 17 = 150 | Mexp = 150.82 | Diff = 0.54 % |
| | | i=10  b= ±1, s= ±1, c = 0 mesons | | |
| *n=0* | B⁰$_s$ | M(BB) = 7x19 + 19 = 152 | Mexp = 153.43 | Diff = 0.93 % |
| | | i=11  1⁻ b= ±1, s=0 mesons | | |
| *n=0* | B* | M(BB) = 7x19 + 23 = 156 | Mexp = 152.14 | Diff = -2.50 % |
| | | i=12  ½ - (b= -1) baryons | | |
| *n=0* | Λ⁰$_b$ | M(ΛΛ) = 7x21 + 13 = 160 | Mexp = 160.68 | Diff = 0.40 % |
| | | i= ?  1⁻ b$\bar{b}$ mesons | | |
| *n=0* | Y(1S) | M(YY) = 7x25 + 95 = 270 | Mexp = 270.28 | Diff = 0.10 % |

## Table IV – Mesons

| (a) Pseudoscalar | $v_i$ | $C_i$ | $m_i$ | h, l, k |
|---|---|---|---|---|
| i=0<br>0⁻ mesons | 1 | 0 | 0 | 1; 0; 0 pion<br>1; 1; 2<br>1; 2; 2 |
| i=4<br>0⁻-charmed mesons<br>(c=1) | 2 | 3 | 1 | 0; 0; 0 |
| i=8<br>0⁻- c$\bar{c}$ mesons | 2 | 4 | 2 | 0; 0; 0 |
| i=9<br>0⁻ b= ±1, s=0 mesons | 3 | 5 | 2 | 1; 2; 3 |
| i=10<br>b= ±1, s= ±1, c = 0 mesons | 3 | 5 | 2 | 1; 3; 3 |

| (b) Pseudovectorial | $v_i$ | $C_i$ | $m_i$ | h, l, k |
|---|---|---|---|---|
| i=1<br>1⁻- mesons | 1 | 1 | 0 | 1; 2; 2<br>1; 2; 3<br>1; 3; 3 |
| i=5<br>1⁻-charmed mesons<br>(c=1) | 2 | 3 | 1 | 0; 1; 1 |
| i=8<br>1⁻ c$\bar{c}$ mesons | 2 | 4 | 2 | 0; 1; 1 |
| i=11<br>1⁻ b= ±1, s=0 mesons | 3 | 5 | 2 | 1; 3; 4 |
| i= ?<br>1⁻ b$\bar{b}$ mesons | 3 | 6 | 3 | 1; 15; 16 |

**Notes to Table IV**

1) Equation (6.2a) applies. $C_0 = 0$.
2) $v_i = 1$ for mesons solely constituted by u, d, s quarks and their antiquarks.
3) $v_i$ increases by 1 for each flavour added in the SU(3) multiplet in question.
4) It can be supposed that for mesons an even value of $v_i$ entails that $h$ is null and, on the contrary, that an odd value of $v_i$ entails that $h = 1$.
5) The "levels" identified by the quantum numbers $l, k$ tend to "fill up" according to a regular sequence $lk$ = 00, 01, 11, 12, 22, 23, 33, 34, etc.
6) If the pion 00 is replaced with a "virtual" quantum 11 one has, for the pseudoscalars $i = 0$ containing only u, d, s quarks, the sequence $lk$ = 11, 12, 22; at this point, the sequence is transferred to the pseudovectorials u, d, s ($i = 1$) giving 22, 23, 33. The same result is reached by adding 11 to the values of $lk$ for the pseudoscalars: 11 becomes 22, 12 becomes 23, 22 becomes 33.
7) This rule is also valid for $c\bar{c}$ mesons and for charmed mesons. For pseudoscalars one has 00 and for pseudovectorials 11. $\eta_C$ and J/$\psi$ are thus linked through this rule.
8) $m_i = 0$ for u, d, s baryons; $m_i = 1$ for charmed baryons. Every time a given flavour is equilibrated by the corresponding antiflavour, $m_i$ value increases of 1 (see also Table II).

## Table V – ½-spin baryons

|  | $v_i$ | $C_i$ | $m_i$ | h, l, k |
|---|---|---|---|---|
| $i=2$ ½-baryons | 1 | 2 | 0 | 1; 1; 2<br>1; 2; 3<br>1; 3; 3<br>1; 3; 4 |
| $i=6$ 1/2-charmed (c=1) baryons | 2 | 4 | 1 | 0; 0; 0<br>1; 0; 1<br>1; 1; 1<br>1; 1; 2 |
| $i=12$ ½ - (b= -1) baryons | 3 | 6 | 2 | 1; 2; 2 |

**Notes to Table V**

1) Equation (6.2b) applies.
2) $v_i = 1$ for baryons with 1/2 spin constituted solely by u, d, s quarks and their antiquarks.
3) $v_i$ increases by 1 for each flavour added in the SU(3) multiplet in question.
4) $m_i$ is progressive.
5) If $k = 0$, then $h = 0$; otherwise $h = 1$.
6) The "filling" procedure of "levels" $l, k$ is the same as in mesons. Yet, strangely, levels c = 1 are filled first, then the c = 0 ones.

**Table VI - 3/2-spin baryons**

|  | $v_i$ | $C_i$ | $m_i$ | h, l, k |
|---|---|---|---|---|
| i=3<br>3/2-baryons | 2 | 3 | 0 | 0; 0; 0<br>0; 0-1; 1<br>0; 1-2; 2<br>0; 2; 3 |
| i=7<br>3/2-charmed<br>(c=1) baryons | 2 | 4 | 1 | 1; 1; 1<br>1; 1; 2 |

**Notes to Table VI**

1) Equations (6.2b), (6.2c) apply.
2) $m_i$ is progressive.
3) It follows that $C_3 = 3$ and therefore $v_3 = 2$.
4) The "filling" procedure of "levels" $l, k$ is the same as in mesons. The c = 0 levels are filled first, then the c = 1 levels.
5) For $i = 3$, $k$ is progressive. The successive filling of this multiplet is regular. It must be noted that $k = 1$ is compatible both with $l = 0$ and with $l = 1$, since $h = 0$. For the same reasons, $k = 2$ is compatible both with $l = 1$ and with $l = 2$.
6) The first level of the multiplet $i = 7$ is obtained by adding the usual "virtual" quantum 11 to the first level of the multiplet $i = 3$, which is 00. The result, of course, is 11.

**Appendix II - Hadronic wave equation**

Let L be the radius of the De Sitter hadronic micro-universe, with $L \leq \hbar c/E_0$, $E_0 = 70$ MeV. The j-th quark is a fermion with ½ spin placed on this micro-universe. Dirac's equation for a fermion on De Sitter space (19) can be written thus ($\mu,\nu = 0,1,2,3$):

$$[i\hbar \Gamma^\mu \partial_\mu - (m' - \frac{2i\hbar}{Lc})c]\psi(\vec{x},t) = 0, \qquad \text{A1}$$

$$\Gamma^\mu = \chi\gamma^\mu + \frac{1}{2L}[\gamma^\mu, \gamma_\nu]x^\nu, \qquad \text{A2}$$

$$\chi = \left(1 + \frac{x_\mu x^\mu}{L^2}\right)^{1/2}, \qquad \text{A3}$$

$$m'^2 = m^2 - \frac{\hbar^2}{4L^2 c^2}. \qquad \text{A4}$$

In equation A4, m is the quark mass and m' is its value corrected for the finite size of the chronotope. We assume that individual quarks do not have a defined mass, so that equation A1 is simplified into:

$$(i\hbar \Gamma^\mu \partial_\mu)\psi(\vec{x},t) = 0. \qquad \text{A5}$$

We are seeking an equation in which only the global skeleton mass of the hadron appears. Let us define the k-th matrix α of the j-th quark as $\alpha^{kj} = (\Gamma^{0j})^{-1}(\Gamma^{kj})$. If we limit ourselves, in the first stage, to considering only electromagnetic interactions, we have:

$$i\hbar \partial_t \Psi = \sum_{j=1}^{N} \left[-\vec{\alpha} \cdot (\vec{\partial} - q\vec{A}) + i\hbar q A_0 \right]_j \Psi + M_{sk} \Psi \qquad A6$$

as a wave equation for the global hadronic wavefunction $\Psi(\vec{x}_1, ..., \vec{x}_j, ..., \vec{x}_N, t)$.

In order to include the colour forces, as well, we consider the expansion:

$$\psi(\vec{x}_j, t) = \sum_\alpha \psi_\alpha u_\alpha ,$$

where the $u_\alpha$ functions constitute a complete orthonormal base on the colour space of the individual quark. The covariant derivation with respect to the colour SU(3) group yields:

$$D_\mu \psi_\beta = \sum_\alpha [\delta_{\beta\alpha} \partial_\mu - iq_s (A_\mu)_{\beta\alpha}] \psi_\alpha = \partial_\mu \psi_\beta - iq_s \sum_\alpha [(A_\mu)_{\beta\alpha}] \psi_\alpha =$$

$$= \partial_\mu \psi_\beta - iq_s (A_\mu)^\alpha_\beta \psi_\alpha .$$

$$D_\mu \psi = \partial_\mu \psi - iq_s (A_\mu)^{\beta\alpha} \psi_\beta u_\alpha .$$

In the expressions above, $q_s$ is the colour charge of the j-th quark. Let:

$$(A_\mu)^{\beta\alpha} \psi_\beta u_\alpha = \hat{A}_\mu \psi , \qquad A7$$

the inclusion of colour interaction in equation A6 leads to equation (2.6).

As regards the electromagnetic potentials $A_\mu$ which appear in equation (2.6), the one experienced by the j-th quark is the solution to the D'Alembert equation on the De Sitter micro-universe:

$$\Box (A_\mu)_j = \frac{-4\pi (J_\mu)_j}{c} , \qquad A8$$

in which the D'Alembert operator is expressed as a function of the coordinates of the j-th quark and of the partial derivatives with respect to those coordinates (20, 21):

$$L^2 \Box = \chi^2 (L^2 \partial_k \partial^k + x^i x^k \partial_i \partial_k + 2 x^i \partial_i) , \qquad A9$$

while the four-current $(J_\mu)_j$ generated by the other quarks – or by the j-th quark itself, if one wishes to consider its self-interaction - is calculated starting from $\Psi$.

In colour potentials $(\hat{A}_\mu)_j$, on the other hand, factors are present whose form (2, 9) is:

$$\prod_{j' \neq j} \left(1 - \frac{d_{jj'}}{d_0}\right)^{-1} , \qquad A10$$

where $d_{jj'}$ is the distance of the j-th quark from the j'-th quark and $d_0 = e^2/mc^2$, with m = mass of the electron.

As a subsidiary condition, only colour singlets will be acceptable hadronic wavefunctions.